# Phase Locked Photon Echoes for Near-Perfect Retrieval Efficiency and Extended Storage Time

Byoung S. Ham[1,2,3,*] and Joonseong Hahn[1,2]

[1]Center for Photon Information Processing, [2]the Graduate School of Information & Telecommunications, and [3]Department of Electrical Engineering, Inha University
253 Yonghyun-dong, Nam-gu, Incheon 402-751, S. Korea
[*]bham@inha.ac.kr

**Abstract:** Quantum storage of light in a collective ensemble of atoms plays an important role in quantum information processing. Consisting of a quantum repeater together with quantum entanglement swapping, quantum memory has been intensively studied recently. Conventional photon echoes have been limited by extremely low retrieval efficiency and short storage time confined by the optical phase decay process. Here, we report a storage time-extended near perfect photon echo protocol using a phase locking method via an auxiliary spin state, where the phase locking acts as a conditional stopper of the rephasing process resulting in extension of storage time determined by the spin dephasing process. We experimentally prove the proposed phase locked photon echo protocol in a $Pr^{3+}$ doped $Y_2SiO_5$ in a backward propagation scheme, where the contradictory dilemma of using optically dense medium in photon echo methods can be solved.

Quantum interface is essential to quantum information processing, where flying qubits interact with an atomic medium[1-7]. The importance of quantum optical data storage varies, from applications of shorter storage time for quantum delay[8,9] in quantum computing to longer storage time of quantum memory[1-7] for quantum repeaters[10,11], which enable long-distance quantum communications. Recent observation of quantum interface using dynamic spectral gratings illustrates a modified version of the conventional photon echoes[12-14] to enhance retrieval efficiency in a dilute sample[1]. Unlike most other quantum memories limited by single mode storage[2-6], photon echoes use the reversible rephasing phenomenon in a collective atomic ensemble utilizing inhomogeneous broadening, so that consecutive multiple bit storage or multimode storage is possible[1,15-22]. From the standpoint of practical quantum memories, thus, the quantum memory should be determined by: 1. How long the quantum optical data is stored; 2. How wide the bandwidth is, and 3. How high retrieval efficiency is obtained. Regarding storage time, several modified photon echo techniques have resulted in storage time extension, replacing a constraint limited by optical dephasing with spin dephasing, which is normally one order of magnitude longer. Examples include controlled reversible inhomogeneous broadening (CRIB) (refs. 15-18) and frequency comb-based spectral grating methods[1,15]. A recent proposal of resonant Raman echoes using optical locking shows a breakthrough in the storage time by extending it to spin population relaxation time[7]. Regarding echo efficiency, several modified methods have been presented so far: 1. A backward propagation scheme using an auxiliary spin state to overcome the geometry-based intrinsically low photon echo efficiency[16-18]; 2. Direct atom swapping by using a DC Stark field for rephasing[18,19]; and 3. Slow light-enhanced photon echoes to lengthen the light-matter interaction time for enhanced absorption, even in a dilute optical medium[21,22]. The slow light-enhanced photon echo observation in particular has demonstrated a few hundred times increased photon echo efficiency[22]. However, most of the modified versions of the photon echoes mentioned above need an additional process, which may be a drawback. Examples include: 1. External DC field usage for the rephasing process[18,19]; 2. A long optical pumping process for frequency comb generation[1,15]; 3. Ratio frequency usage for rephasing[17,20]; and 4. Requirement of electromagnetically induced transparency[23], which is difficult to obtain in most solid media[21].

In this Article we report phase locked photon echoes to extend the photon storage time as well as to obtain near perfect retrieval efficiency. The storage time extension results from phase locking via an auxiliary spin state, where the phase locking temporally freezes the optical rephasing process by transferring the excited atoms to an auxiliary spin state. Thus, the storage time is now determined by the spin dephasing process of the auxiliary spin state. The distinctive difference of the present method from the CRIB is the optical control of the photon echo rephasing process without relying on external DC field and/or rf pulses. The backward propagation photon echo scheme has been originally introduced in conventional three-pulse photon echoes[13,14]; otherwise we meet a contradictory dilemma between data photon absorption and echo reabsorption[22,24]. Moreover, the backward propagation photon echoes have potential for aberration corrections when dealing with quantum imaging applications[25]. Regarding the controversial point of π rephasing pulse-induced spontaneous noise in conventional two-pulse photon echoes[26], the present technique avoids this controversy with use of a spectrally and spatially independent π rephasing pulse. Thus, the present phase locked photon echo protocol contains all the benefits of photon echoes with the additional advantages of storage time extension up to the spin phase decay time, which is one order of magnitude longer and near perfect retrieval efficiency due to the backward propagation scheme.

Figure 1 shows a theoretical demonstration of the present phase locked photon echo method for storage time extension with near perfect retrieval efficiency. In Fig. 1a the states |1> and |3> interacting with quantum field or weak classical light (D, data) satisfy the conventional photon echo scheme. The light B1 and B2 transient with an auxiliary spin state |2> are used for phase locking and unlocking of the photon echo process initiated by R (rephasing). Here the phase locking means a controlled stoppage of the rephasing process via a complete transfer of excited atoms on state |3> to state |2> using a π pulse area of B1. Figure 1b represents the pulse sequence of the present phase locked photon echo protocol. The first two red bars, D and R, configure a conventional photon echo sequence. The pulse area of R should be π for complete rephasing. The blue bars, B1 and B2, respectively, play phase locking and unlocking by temporally stopping and resuming the rephasing process initiated by R. As a result, photon echo E follows at:

$$T_E = T_{B2} + (T_R - T_D) - (T_{B1} - T_R). \tag{1}$$

Figure 1c represents a programmed pulse sequence in the numerical calculations, and results are shown in Fig. 1d. To maximize the photon echo efficiency, D is set to be π/2. However, the echo efficiency is independent of the D pulse area (see Supplementary Fig. S1a). Quantum field treatment of using very weak coherent light has already been discussed[27]. The numerical simulation is shown in Supplementary Fig. S1b, and is experimentally demonstrated in Fig. 4b. In Fig. 1c, however, we neglect the delay of B1 $[(T_{B1} - T_R) \ll T_R]$ to directly compare the phase locked photon echoes with conventional photon echoes, so that $T_E - T_{B2}$ is the same as $T_R - T_D$ in equation (1), which positioned at t=60 μs.

For the numerical calculations, nine time-dependent density matrix equations are numerically solved without any assumption under the rotating wave approximation. The density matrix approach is very powerful in dealing with an ensemble system interacting with coherent laser fields owing to statistical information as well as quantum mechanical information[28]. The equation of motion of the density matrix operator $\rho$ is determined from Schrödinger's equation[28]:

$$\frac{d\rho}{dt} = -\frac{i}{\hbar}[H,\rho] - \frac{1}{2}\{\Gamma,\rho\}, \tag{2}$$

where $\{\Gamma,\rho\}$ is $\Gamma\rho + \rho\Gamma$, $H$ is the Hamiltonian, $\hbar$ is the Planck's constant divided by $2\pi$, and $\Gamma$ is decay rate. The density operator $\rho$ is defined by $\rho = |\Psi\rangle\langle\Psi|$, where $|\Psi\rangle$ is the state vector. By solving equation (2) the following are obtained as main coupled equations:

$$\frac{d\rho_{13}}{dt} = i\Omega_{13}(\rho_{33} - \rho_{11}) - i\Omega_{23}\rho_{12} - i(\delta_1 + \gamma_{13})\rho_{13}, \tag{3}$$

$$\frac{d\rho_{23}}{dt} = i\Omega_{23}(\rho_{33} - \rho_{22}) - i\Omega_{13}\rho_{21} - i(\delta_1 + \gamma_{23})\rho_{23}, \tag{4}$$

$$\frac{d\rho_{12}}{dt} = i\Omega_{13}\rho_{32} - i\Omega_{23}\rho_{13} - i(\delta_1 - \delta_2)\rho_{12} - \gamma_{12}\rho_{12}, \tag{5}$$

where $\Omega_{13}$ ($\Omega_{23}$) is D or R (B1 or B2) Rabi frequency, and $\gamma_{ij}$ is the phase decay rate between states $|i\rangle$ and $|j\rangle$ (see Fig. 1a). For simplicity, spin transition is assumed to be homogeneous, and both optical transitions are inhomogeneously broadened by $\Delta_{inh}$=680 kHz (full width at half maximum, Gaussian distribution). The optical inhomogeneous broadening $\Delta_{inh}$ is equally divided by a 10 kHz space for 161 divisions in the calculations. The results are the same for weak or strong field of D. For visual effect, we use maximum coherence excitation by D, which is $\pi/2$ pulse area. Rabi frequency of all pulses is set at 5 MHz.

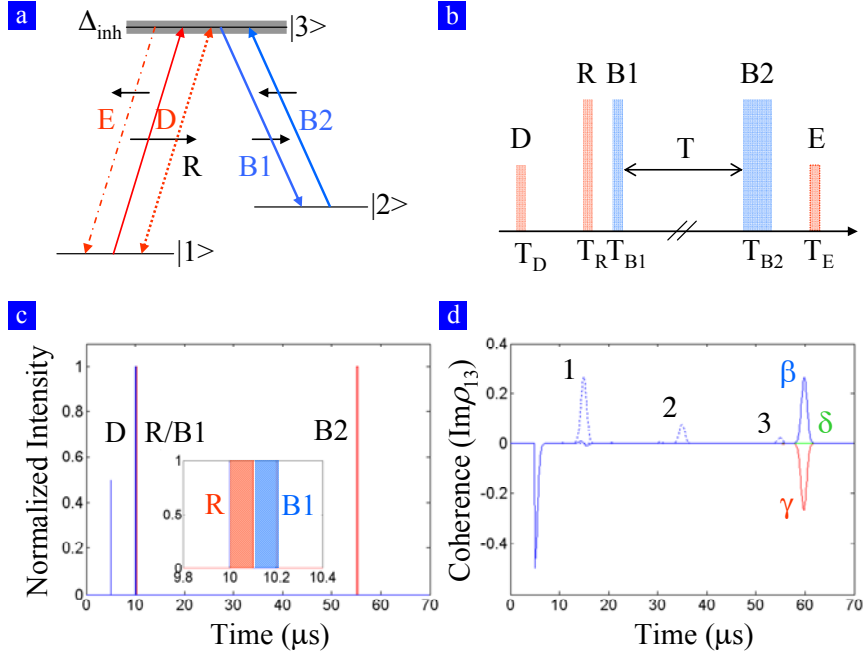

**Figure 1 Phase locked echo. a,** Energy level diagram interacting with light pulses. B1, B2, D, E, and R stand for locking, unlocking, data, photon echo, and rephasing pulses. The pulse areas of D, R, B1, and B2 are $\pi/2$, $\pi$, $\pi$, and $3\pi$, respectively. $\Delta_{inh}$ (680 kHz at full width at half maximum) is an optical inhomogeneous width with Gaussian distribution. **b,** Pulse sequence. **c,** Programed pulse sequence. $T_D$=5; $T_R$=10; $T_{B1}$=10.1; $T_{B2}$=55 μs. **d,** Numerical simulations for (c). Marks 1, 2, and 3 represent conventional photon echoes without B1 and B2 for $T_R$=10, 20, and 30 μs, respectively. Marks β, δ, and γ respectively represent phase locked photon echoes with B1 ($\pi$) and B2 for B2=$3\pi$, $2\pi$, and $\pi$ in pulse area. The pulse area is defined by $\int \Omega dt$. $\Gamma_{12}=\gamma_{12}=0$; $\Gamma_{13}=\Gamma_{23}=5$ kHz; $\gamma_{13}=\gamma_{23}=10$ kHz. $\Omega_D=\Omega_R=\Omega_{B1}=\Omega_{B2}=5$ MHz.

Figure 1d shows numerically calculated results of the present phase locked photon echoes. The curves with marks "1," "2," and "3" represent conventional photon echoes without B1 and B2, where the echo efficiency exponentially drops according to the optical phase decay time $T_2^{Opt}$ (8 μs) (see also Fig. 4d for experimental results). In contrast, the photon echo efficiency of β with B1 and B2 shows the same value as the conventional one marked by "1," if the pulse area of B2 satisfies $3\pi$. The storage time extension by B2 is determined by spin dephasing between states $|1\rangle$ and $|2\rangle$, which is assumed zero in Fig. 1d. If the pulse area of B2 does not exceed $2\pi$, photon echo generation cannot be expected (see δ for 0% with $2\pi$ of B2 or γ for −100% with $\pi$ of B2 as will be discussed in Fig. 3). For proper phase locking, B1 must be applied before the rephasing process completes (see Supplementary Fig. S2). Although the optical phase is locked by B1, the transferred atoms should decay according to the spin dephasing (see

Supplementary Fig. S3). Here we emphasize that the R-induced spontaneous noise in conventional two-pulse photon echoes is completely eliminated in the present phase locked scheme through coherent population transfer from the excited state to the third spin state by B1. We also note usage of different frequency and different propagation direction of B1 and B2 compared with that of D.

To achieve near 100% retrieval efficiency, however, a forward scheme of photon echo must be avoided due to reabsorption of the photon echoes by the residual atoms. This has been the main contradictory dilemma of photon echoes, because an optically dense medium is required for complete data absorption, while generated photon echoes experience more absorption as they propagate a longer distance due to the exponential decay of the data pulse absorption[24]. To solve this dilemma, a backward propagation scheme has been introduced in a modified photon echo scheme for nearly 100% echo efficiency[16-18]. Thus, complete photon echo recovery without reabsorption can be obtained. The backward propagation scheme acts as a phase conjugate scheme, which is greatly beneficial, when dealing with quantum images because spatial aberration cancellation is expected[25,29,30].

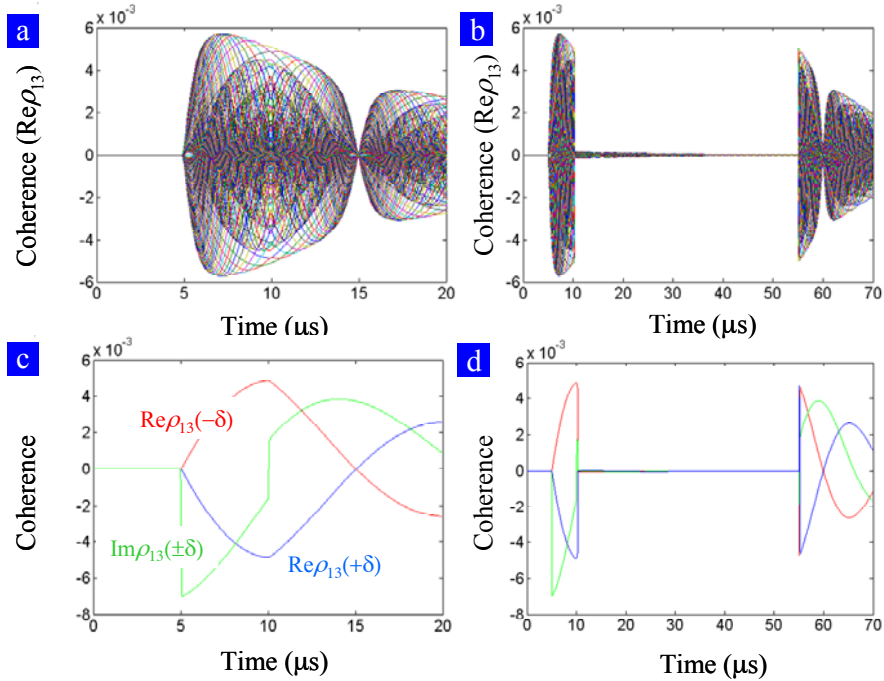

**Figure 2 Storage time extension by B1. a** and **c,** Conventional two-pulse photon echo. D (R) turns on at t=5 (10) μs. δ=40 kHz. **b** and **d,** Phase locked photon echo using B1 (π) and B2 (3π). 161 groups of atoms are considered for 1.6 MHz optical inhomogeneous broadening. All parameters are the same as in Fig. 1.

Figure 2 shows analysis of the phase locking process presented in Fig. 1 in comparison with conventional two-pulse photon echoes. Figures 2a and 2c represent individual atom phase evolution. Each atom's phase velocity is determined by the detuning δ in the optical inhomogeneous broadening $\Delta_{inh}$. After the data pulse excitation, sum phases of all atoms decay out quickly, but the initial coherence is recovered by the rephasing pulse R. Figure 2c is for a symmetrically detuned atom pair (red and blue curves) to depict atom phase swapping by R at t=10 μs. This atom phase swapping is a fundamental characteristic to the time reversed process in photon echoes. Figures 2b and 2d show the evolution of phase locked atoms by B1 and B2 (for β in Fig. 1d), revealing exactly the same features as in Figs. 2a and 2c, except for the storage time extension during the period between B1 and B2. This storage time extension is determined by the spin dephasing time of state |2>, which is set to zero in Fig. 2, for direct comparison.

Figure 3 presents details of how the phase locking and unlocking play, as presented in Fig. 2d. Figures 3a and 3b represent population density of each state, while Figs. 3c and 3d represent corresponding coherence, respectively. For the rephasing R (pink region), the population swapping between states |1> and |3> by the π R pulse (see Fig. 3a) induces phase swapping between symmetrically detuned atoms by ±δ as discussed in Fig. 2c. The π pulse of B1 induces population swapping between states |2> and |3> (see blue region of Fig. 3a), resulting in an additional π/2 phase shift between the symmetrically detuned atoms (±δ) (see blue region of Fig. 3c). This π/2 phase shift ends up at zero coherence resulting in freezing of phase evolution. This phase locking continues until the phase unlocking pulse B2 arrives. Because the population swapping between states |2> and |3> induces π/2 phase shift, it needs an additional 3π/2 phase shift to achieve the same phase recovery (see Fig. 2d), accomplished by the rephasing pulse R. This is the key mechanism of the present method of phase locking. Unlike rephasing pulse R, the phase locking and unlocking pulses need twice the energy to shift the same amount of phase. This is because the pulse of B1 or B2 interacts with only state |3>, while R interacts with both states |1> and |3>. More detailed discussions appear below and in Supplementary Fig. S4.

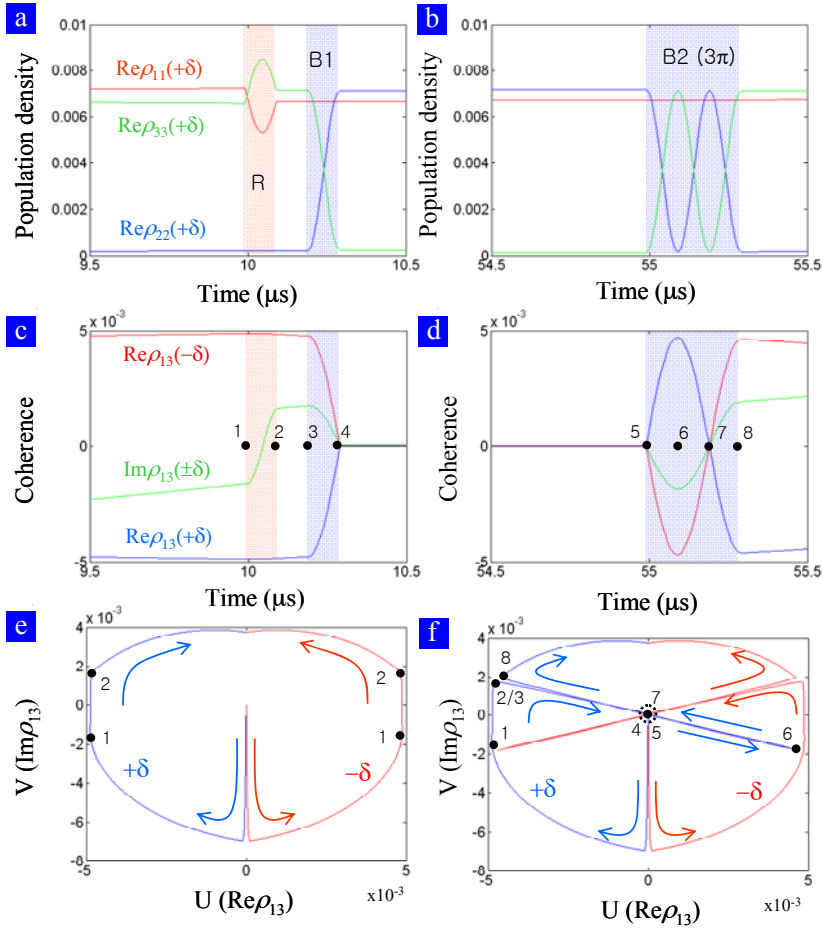

**Figure 3 Rephasing control by using B pulses. a** and **b,** Population evolution with R, B1, and B2 for symmetrically detuned atoms. **c** and **d,** Coherence evolution with R, B1, and B2 for symmetrically detuned atoms. **e** and **f,** Bloch vector evolution for (c) and (d), respectively. All parameters are the same as in Fig. 2, unless otherwise indicated.

Figures 3e and 3f depict Bloch vector (u, v) evolution in the uv plane for symmetrically detuned atoms by δ without (Fig. 3e) and with (Fig. 3f) phase locking and subsequent unlocking by B1 and B2. Dots with numbers denote timing of the pulse. As discussed in Fig. 2, the rephasing pulse R swaps symmetrically detuned atoms, so that the δ detuned atom (blue curve) after dot 2 must follow the −δ detuned atom trajectory (red curve) in time reversed manner as shown in Fig. 3e. In Fig. 3f, the phase locking pulse B1 turned on at dot 3 (for +δ) renders the Bloch vector shrunk to zero reaching at dot 4, meaning zero coherence magnitude in both absorption (Im$\rho_{13}$) and dispersion (Re$\rho_{13}$). Thus the phase evolution is locked until B2 arrives. The B2 pulse pumps atoms from state |2> back to state |3>, resulting in absorption increase until complete depletion occurs by the π pulse ending at dot 6. Dot 6 is for γ in Fig. 1d leading to −100 % photon echo efficiency because it must follow the red curve in Fig. 3f. Thus, to return back to the same position as dot 3, additional 2π pulse area is needed. This proves why B2 must be 3π. Thus, we can conclude that the pulse area of R, B1, and B2 must satisfy following equations for maximum photon echoes:

$$\Phi_R = (2n-1)\pi, \quad (6)$$

$$\Phi_{B2} = (4n-1)\pi \text{ for } \Phi_{B1} = (4n-3)\pi, \quad (7)$$

$$\Phi_{B2} = (4n-3)\pi \text{ for } \Phi_{B1} = (4n-1)\pi, \quad (8)$$

$$\Phi_{B1+B2} = 4n\pi, \quad (9)$$

where $\Phi_i$ is the pulse area of pulse $i$, and n is an integer (see Supplementary Fig. S4).

Figure 4 shows experimental proofs of the proposed phase locked photon echo protocol. The experiment was performed in a rare-earth $Pr^{3+}$ (0.05 at. %) doped $Y_2SiO_5$ cooled to 5K. The sample's length is 1 mm, and spectral modification is performed using repumping processes to increase photon absorption[22]. The light pulses in Fig. 4a are from a ring-dye laser (Tecknoscan) via acoustooptic modulators (Isomet) driven by digital delay generators (Stanford DG 535) and radiofrequency synthesizer (PTS 250). The angle among D and B1 is ~12.5 mili-radians and overlapped by 90% through the sample of 1 mm in length. The spot diameter of B1, B2, and D & R is 330, 200, and 300 μm, respectively. The pulses D and R are from the same laser output modulated by an acoustooptic modulator. The light power of R & D, B1, and B2 is 0.5, 12, and 24 mW. The optimum power of the light B1 and B2 in Fig. 4 is predetermined by Rabi flopping measurement, where B1 (B2) pulse area is set at π (3π). The optical signals detected by an avalanche photodiode are recorded in the oscilloscope by averaging 30 samples. The repetition rate of the light pulse sequence is 20 Hz. The temperature of the sample is kept at ~5 K. To utilize one (±5/2, $^3H_4$) of the three ground states as an auxiliary spin state for B1 and B2, the state (±5/2, $^3H_4$) is pre-pumped out to vacate before sending a data pulse D. The effective atom broadening is determined by laser jitter which is ~300 kHz. The light is vertically polarized and the crystal axis of the medium, Pr:YSO is along the propagation direction **k**. The temperature of the sample of Pr:YSO is controlled by the liquid helium flow rate in a cryostat (Advanced Research System). A perfect phase conjugate scheme, however, does not work due to a frequency difference between D/R and B1/B2. The resultant absorption of a data pulse D (1 μs) is ~70%. Figure 4a shows the pulse propagation scheme of Fig. 1a. According to the phase matching condition, the echo signal E is backward:

$$\vec{k}_E = \vec{k}_D - \vec{k}_{B1} + \vec{k}_{B2}, \quad (10)$$

$$\varpi_E = \varpi_D - \varpi_{B1} + \varpi_{B2}, \quad (11)$$

where $\vec{k}_i$ and $\varpi_i$ are wave vector and angular frequency of the pulse $i$, respectively. Equation (10) denotes a coherence transfer-based four-wave mixing which resulted from spin coherence $\rho_{12}$ excitation by B1 and coherence readout into an echo E by B2. Unlike resonant Raman coherence excitation[7,21], the spin coherence $\rho_{12}$ is created via optical population transfer from states |3> to |2> by B1, so that the four-wave mixing signal as an echo depends on the rephasing process. In other words, the Raman induced spin coherence excited by R and B1 is negligible because they are temporally separated not to keep optical coherence: The optical free induction decay time is as short as 1 μs. Although E is not a phase conjugate of D, it actually plays as a phase conjugate if the spatial shape of B1 and B2 is properly controlled for correlation and convolution theorem[13]. Conventional two-pulse photon echo is detected by the APD1, and the phase locked photon echo is detected by the APD2.

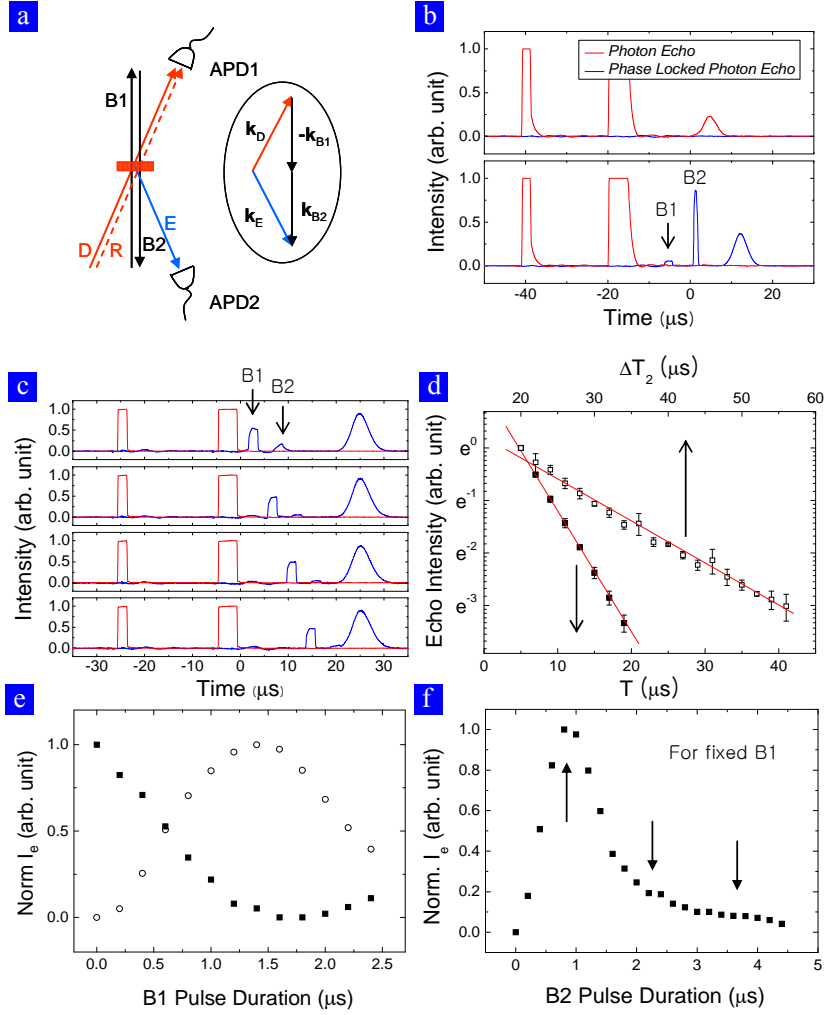

**Figure 4 Experimental results of the phase locked photon echoes. a,** Backward propagation scheme for a quasi phase conjugate. **b,** (upper panel) Conventional two-pulse photon echo; (lower panel) phase locked photon echo. **c,** B1 position invariant phase locked echoes. **d,** Dephasing time of conventional photon echo ($\Delta T_2$), and phase locked photon echo (T). **e,** Collective coherence transfer; closed squares: photon echo (red echo in b); open circles: phase locked echo (blue echo in b). **f,** B2 dependent phase locked echo. In **b** and **c**, red (blue) line is from APD1 (APD2). In blue line, B1 and B2 are scattered light to denote their positions.

The upper panel of Fig. 4b represents the conventional two-pulse photon echo (without B1 and B2) as a reference, with an observed echo efficiency (echo intensity ratio to data intensity) of ~ 0.3%. The lower panel of Fig. 4b presents the phase locked photon echo in Fig. 1, where T (B2 delay from B1) is 5 µs and ($T_R - T_D$) is 20 µs. The delay of B1 from R is 7 µs. The calculated delay of echo E from B2 by equation (1) matches the observed delay at 13 µs as shown in Fig. 4b (see also Supplementary Fig. S1b). The auxiliary state |2> of Pr:YSO is not isolated but closely connected to state |1> with a spin dephasing time of 9 µs, determined by the spin inhomogeneous broadening (35 kHz) (to be discussed in Fig. 4d). This means that delay T is limited by 10 µs. Although the echo is not a phase conjugate of the data D, higher retrieval efficiency of the absorbed D as an echo is still achieved owing to phase conjugate-like backward propagation. In Fig. 4b the phase locked echo intensity is twice that of the conventional two-pulse photon echo. Considering coherence loss of 67% [$\exp(-5/9)^2$; will be discussed in Fig. 4] during B1 and B2 due to the spin dephasing between states |1> and |2>, we conclude that the enhancement factor 6 is

achieved due to backward propagation of E, which traces back the same path as the data pulse does, so that residual atom absorption caused loss is negligible, as proposed in refs. 16 and 17. Regarding storage time extension, Fig. 4b shows proof of principle in the limit of fast spin dephasing time of 10 μs. However, modification of the spin dephasing to extend storage time can be achieved simply by applying an external magnetic field gradient[31]. Ultimately, if the auxiliary state |2> is isolated from state |1>, the storage time can be extended up to spin population decay time, which is minutes.

Figure 4c proves equation (1), where the echo position $T_E$ is invariant of the delay B1 from R for a fixed T. As discussed in Supplementary Fig. S1b, the experimental results in Fig. 4c prove that the delay B1 does not change the echo intensity. This is because the function of B1 and B2, respectively, is to lock and unlock the rephasing process triggered by R. With the delay T bigger than $(T_R-T_D)$, we observed no phase locked echo signal, but a conventional photon echo appeared as in the upper panel of Fig. 4b (see also Supplementary Fig. S2). In Fig. 4d the photon echo intensities are measured as functions of delay of R (upper axis, $\Delta T_2$) and B2 (lower axis, T) for a fixed B1. We calculated decay time τ for the R delay at τ =25μs, and for the B2 delay at τ =9μs: $I_e(t)=I_e(0)exp(-2t/\tau)$, where $I_e$ is intensity of echoes. The B2 delay dependent decay time is analogous with the inverse of the spin inhomogeneous width ($\Delta_{spin}=1/\pi\tau=30$ kHz) between states |1> and |2>. The R delay dependent decay time is analogous to the inverse of optical homogeneous decay time, where the optical homogeneous decay time shortens rapidly as temperature increases (see Fig. 4b of ref. 32).

Figure 4e shows collective coherence transfer in the present phase locked scheme. The conventional two-pulse photon echoes (filled squares; see also red echo in Fig. 4b) are transferred into phase locked echoes (open circles; see blue echo in Fig. 4b) as B1 pulse duration ΔT increases for a fixed B2. At ΔT=1.6μs, the two-pulse photon echo is transferred completely into the phase locked echo, which denotes maximum population transfer from state |3> to state |2> with a π pulse area of B1 (see Fig. 3a). Figure 4f proves equation (8). For a fixed pulse area of B1 (1.6 μs, π pulse area), the first maximum echo intensity appears at 0.9 μs of B2 pulse duration, and the second and third maxima appear at ~2.1 μs and ~3.6 μs, respectively: These pulse durations roughly correspond to 3π, 7π, and 11π, respectively, to satisfy equation (5). These echoes, however, turn out to be saturated due to long pulse length of B2 compared with the ~10 μs spin dephasing time discussed in Fig. 4d. For the π pulse area of B1, the pulse area of B2 must satisfy 3π (0.9 μs), 7π (2.1 μs), and 11π (3.3 μs) for the maximum position according to equation (9). Thus, Fig. 4 experimentally proves the theoretical analyses discussed in Figs. 1~3, and the previously proposed modified photon echo protocols (refs. 16-18) must be limited by the same constraint of spin dephasing time.

In conclusion, a phase locked photon echo protocol was proposed and experimentally proved in a rare-earth $Pr^{3+}$ doped $Y_2SiO_5$, where conventional two-pulse photon echo storage time, limited by the optical phase decay process, is extended to spin dephasing time, which is normally an order of magnitude longer in most rare-earth doped solids. The observed photon echo efficiency in a backward propagation scheme was enhanced by a factor of six even in a dilute sample. Thus, the contradictory dilemma of using an optically dense medium is no longer a problem. However, the storage time extension was small due to a short spin dephasing time determined by spin inhomogeneous broadening. Extremely longer storage time with near perfect retrieval efficiency can be obtained if magnetic field gradient is applied or the auxiliary spin state is isolated from the ground state. Quantum imaging with aberration corrections is also another benefit of the present phase locked photon echo protocol based on the correlation and convolution theorem.

**Acknowledgments**
This work was supported by the CRI program (Center for Photon Information Processing) of the Korean government (MEST) via National Research Foundation, S. Korea.

# References

1. de Riedmatten, H., Afzelius, M., Staudt, M. U., Simon, C., & Gisin, N. A solid-state light-matter interface at the single-photon level. *Nature* **456**, 773-777 (2008).
2. Choi, K. S., Deng, H., Laurat, J., & Kimble, H. J. Mapping photonic entanglement into and out of a quantum memory. *Nature* **452**, 67-71 (2008).
3. Honda, K., Akamatsu, D., Arikawa, M., Yokoi, Y., Akiba, K., Nagatsuka, S., Tanimura, T., Furusawa, A. & Kozuma. M. Storage and retrieval of a squeezed vacuum. *Phys. Rev. Lett.* **100**, 093601 (2008)
4. Eisaman, M. D., Andre, A., Massou, F., Fleischhauer, M., Zibro, A. S., & Lukin, M. D. Electromagnetically induced transparency with tunable single-photon pulses. *Nature* **438**, 837-841 (2005).
5. Chaneliere, T., Matsukevich, D. N., Jenkins, S. D., Lan, S.-Y., Kennedy, T. A. B. and Kuzmich, A. Storage and retrieval of single photons transmitted between remote quantum memories. *Nature* **438**, 833-836 (2005).
6. Julsgaard, B., Sherson, J., Cirac, J. I., Fiurasek, J., & Polzik, E. S. Experimental demonstration of quantum memory for light. *Nature* **432**, 482-486 (2004).
7. Ham, B. S. Ultralong quantum optical storage using reversible inhomogeneous spin ensembles," *Nature Photon.* **3**, 518-522 (2009).
8. Zukowski, M., Zeilinger, A., Horne, M. A., & Ekert, A. Event-ready-detectors Bell experiment via entanglement swapping. *Phys. Rev. Lett.* **71**, 4287-4290 (1993).
9. Marino, A. M., Pooser, R. C., Boyer, V., & Lett, P. D. Tunable delay of Einstein-Podolsky-Rosen entanglement. *Nature* **457**, 859-862 (2009).
10. Duan, L.-M., Lukin, M. D., Cirac, J. I., & Zoller, P. Long-distance quantum communication with atomic ensembles and linear optics. *Nature* **414**, 413-418 (2001).
11. Simon, C., de Riedmatten, H., Afzelius, M., Sangouard, N., Zbinden, H., and Gisin, N. Quantum repeaters with photon pair sources and multimode memories. *Phys. Rev. Lett.* **98**, 190503 (2007).
12. Kurnit, N. A., Abella, I. D., and Hartmann, S. R. Observation of a photon Echo. *Phys. Rev. Lett.* **13**, 567-570 (1964).
13. Shen, X. A. & Kachur, R. High-speed pattern recognition by using stimulated echoes. *Opt. Lett.* **12**, 593-595 (1987).
14. Mossberg, T. W. Time domain frequency-selective optical data storage. *Opt. Lett.* **7**, 77 (1982).
15. Afzelius, M., Simon, C., de Riedmatten, H., & Gisin, N. Multimode quantum memory based on atomic frequency combs. *Phys. Rev. A* **79**, 052329 (2009).
16. Moiseev, S. A. & Kroll, S. Complete reconstruction of the quantum state of a single-photon wave packet absorbed by a Doppler-Broadened transition. *Phys. Rev. Lett.* **87**, 173601 (2001).
17. Moiseev, S. A., Tarasov, V. F., & Ham, B. S. Quantum memory photon echo-like techniques in solids. *J. Opt. B: Quantum Semiclass. Opt.* **5**, S497-S502 (2003).
18. Nilsson, M. & Kroll, S. Solid state quantum memory using complete absorption and re-emission of photons by tailored and externally controlled inhomogeneous absorption profiles. *Opt. Commun.* **247**, 393-403 (2005).
19. Hetet, G., Longdell, J. J., Alexander, A. L., Lam, P. K., & Sellars, M. J. Electro-Optic quantum memory for light using two-level atoms. *Phys. Rev. Lett.* **100**, 023601 (2008).
20. Longdell, J. J., Fraval, E., Sellars, M. J., & Manson, N. B. Stopped light with storage times greater than one second using electromagnetically induced transparency in a solid. *Phys. Rev. Lett.* **95**, 063601, (2005). This demonstration is another version of ref. (19), where rf pulses are used for the rephasing process.
21. Turukhin, A. V., Sudarshanam, V. S., Shahriar, M. S., Musser, J. A., Ham, B. S., & Hemmer, P. R. Observation of ultraslow stored light pulses in a solid. *Phys. Rev. Lett.* **88**, 023602 (2001).
22. Hahn, J. & Ham, B. S. Slow light enhanced photon echoes. arXiv:0909.4992.
23. Fleischhauer, M., Imamoglu, A. & Marangos, J. P. Electromagnetically induced transparency: Optics in coherence media. *Rev. Mod. Phys.* **77**, 633-673 (2005).
24. Sangouard, N., Simon, C., Afzelius, M. & Gisin, N. Analysis of a quantum memory for photons based on controlled reversible inhomogeneous broadening. *Phys. Rev. A* **75**, 032327 (2007).
25. Vasilyev, D. V., Sokolov, I. V., & Polzik, E. S. Quantum memory of images: A quantum hologram. *Phys. Rev. A* **77**, 020302(R) (2008).
26. Ruggiero, J., Le Gouet, J.-L., Simon, C., & Chaneliere, T. Why the two-pulse photon echo is not a good quantum memory protocol. Phys. Rev. A **79**, 053851 (2009).
27. Novikova, I., Philips, N. B. & Gorshkov, A. V. Optimal light storage with full pulse-shape control. *Phys. Rev. A* **78**, 021802 (2008).
28. Sargent III, M., Scully, M. O. & Lamb, Jr, W. E. Laser Physics 79–95 (Addison-Wesley, 1974).
29. Boyer, V. Marino, A. M., Pooser, R. C., & Lett. P. D. Entangled images from four-wave mixing. *Science* **321**,

544-547 (2008).
30. Bonato, C. Sergienko, A. V., Saleh, B. E. A., Bonora, S. & Villoresi, P. Even-order aberration cancellation in quantum interferometry. *Phys. Rev. Lett.* **101**, 233603 (2008).
31. Fraval, E., Sellars, M. J., & Longell, J. J. Method of extending hyperfine coherence times in $Pr^{3+}$:$Y_2SiO_5$. Phys. Rev. Lett. **92**, 077601 (2004).
32. Ham, B. S., Shahriar, M. S., Kim, M. K., & Hemmer, P. R. Frequency-selective time-domain optical data storage by electromagnetically induced transparency in a rare-earth-doped solid. *Opt. Lett.* **22**, 1849-1851 (1997).